\documentclass[aps,pra,twocolumn,a4paper,groupedaddress,floatfix,10pt,longbibliography]{revtex4-1}
\usepackage{graphicx,amsmath,amssymb,amsfonts,dsfont,enumitem,MnSymbol}

\usepackage{etoolbox}
\patchcmd{\thebibliography}{\section*{\refname}}{}{}{}

\usepackage{tikz,tikz-3dplot}    
\usetikzlibrary{quantikz}
\usepackage{svg}
\usepackage{amsmath}
\usepackage{multirow}
\usepackage{tikz,tikz-3dplot}    
 \usepackage{amsthm}
\usepackage{hyperref}

\usepackage{blindtext}

\usepackage{color}
\usepackage[caption=false]{subfig}
\newtheorem{lemma}{Lemma}
\usepackage{float}
\RequirePackage{hyperref}
\usepackage{subfiles}

\usepackage{amsmath}

\begin{document}
\title{Symplectic Optimization on Bosonic  Gaussian
States}

\author{Christopher Willby${}^{1,2}$}
\author{Tomohiro Hashizume${}^{1,3}$}
\author{Jason Crain${}^{4,2}$}
\author{Dieter Jaksch${}^{1,2,3}$}

%
\affiliation{${}^1$Institut f{\"u}r Quantenphysik, Universit{\"a}t Hamburg, 22761 Hamburg, Germany}
\affiliation{${}^2$Clarendon Laboratory, University of Oxford, Parks Road, Oxford OX1
3PU, United Kingdom}
\affiliation{${}^3$The Hamburg Centre for Ultrafast Imaging, Luruper Chaussee 149, D-22761 Hamburg, Germany}
\affiliation{${}^4$IBM Research Europe, The Hartree Centre STFC Laboratory,
Sci-Tech Daresbury, Warrington WA4 4AD, UK}

\begin{abstract}
Computing bosonic Gaussian ground-states via variational optimization is challenging because the covariance matrices must satisfy the uncertainty principle, rendering constrained or Riemannian optimization costly, delicate, and thus difficult to scale, particularly in large and inhomogeneous systems. We introduce a symplectic optimization framework that addresses this challenge by parameterizing covariance matrices directly as positive-definite symplectic matrices using unit-triangular factorizations. This approach enforces all physical constraints exactly, yielding a globally unconstrained variational formulation of the bosonic ground-state problem. The unconstrained structure also naturally supports solution reuse across nearby Hamiltonians: warm-starting from previously optimized covariance matrices substantially reduces the number of optimization steps required for convergence in families of related configurations, as encountered in crystal lattices, molecular systems, and fluids. We demonstrate the method on weakly dipole-coupled lattices, recovering ground-state energies, covariance matrices, and spectral gaps accurately. The framework further provides a foundation for large-scale approximate treatments of weakly non-quadratic interactions and offers potential scaling advantages through tensor-network enhancements.

\end{abstract}

\maketitle 
\section{Introduction}
Quadratic bosonic Hamiltonians  form a fundamental class of models with Gaussian ground-states \cite{schuch2006quantum, weedbrook2012gaussian, serafini2017quantum}. These systems admit efficient phase-space representations, enabling the polynomial-time extraction of observables and correlation functions. Unlike fermionic systems, where diagonalization of the coupling matrix alone fully determines the ground-state properties \cite{van1980note, colpa1979diagonalisation}, the bosonic case is governed by an underlying symplectic phase-space structure imposed by the uncertainty principle. As a result, ground-state determination is equivalent to a symplectic diagonalization problem \cite{schuch2006quantum, maldonado1993bogoliubov, colpa1978diagonalization, serafini2017quantum, colpa1986diagonalization}.

This additional structure motivates the development of alternative optimization frameworks, particularly for settings involving finding ground-states for large families of related Hamiltonians where previously computed solutions may be reused.
In addition, the polynomial computational cost associated with symplectic diagonalization places severe practical limits on accessible system sizes. Nevertheless, the ability to compute ground-states of large ensembles of quantum harmonic systems remains of practical and scientific interest, as such models underpin the established many-body dispersion method~\cite{tkatchenko2012accurate,khabibrakhmanov2025accurate}, where access to larger system sizes is of particular importance for understanding the role of collective quantum effects in mesoscale processes, such as protein folding~\cite{stohr2019quantum,gori2023second}.

Variational algorithms provide an alternative to exact diagonalization for solving many-body problems, often trading full spectral information for scalable, problem structure-exploiting optimization over a reduced variational manifold.
 In particular, optimization over the manifold of Gaussian states using Riemannian methods has been explored~\cite{windt2021local,son2021computing,gao2021riemannian,francca2021optimization,gori2025multipartite}. However, these approaches face scalability challenges due to non-trivial constraint handling and the implementation of exponential maps. In particular, the retraction maps used in Ref.~\cite{windt2021local} and the quasi-geodesic approach of Ref.~\cite{gao2021riemannian} require matrix exponentiation, leading to truncated power-series expansions and associated approximation errors. Additionally, the Cayley-transform-based methods employed in Refs.~\cite{son2021computing,gao2021riemannian} require matrix inversions at each gradient evaluation, which become computationally costly as the number of variational parameters increases. 

Our key conceptual advance is a globally unconstrained symplectic parameterization of the Gaussian variational manifold. By constructing covariance matrices directly as positive-definite symplectic matrices via unit triangular factorization, the uncertainty principle is enforced exactly at every optimization step. This eliminates the need for penalty methods, or exponential maps, and converts the Gaussian ground-state problem into an unconstrained matrix trace minimization with exact symplectic structure preserved throughout. 

Beyond single-instance optimization, solutions obtained for one Hamiltonian can be reused to initialize nearby problems. We demonstrate that this warm-start capability enables accurate propagation of ground-state solutions across relative density and geometry sweeps using only a small number of optimization updates per parameter value. Consequently, symplectic optimization is naturally suited to settings involving large families of closely related configurations, such as density- or geometry-dependent scans in molecular fluids and complex chemical environments, where ground-states must be recomputed repeatedly across similar instances \cite{reiher2017elucidating}.

We numerically validate our algorithm by solving ground-states of model quantum chemical systems consisting of ensembles of dipolar-interacting quantum Drude oscillators \cite{cipcigan2019electronic,khabibrakhmanov2025accurate,goger2023optimized} on lattices, demonstrating clear accuracy for non-trivial chemical problems. While lattice implementations serve as convenient benchmarks, the framework itself is not restricted to lattice geometries and is directly applicable to describing quantum dipolar fluctuations in molecular fluids~\cite{jones2013electronically}, where such effects are particularly important at complex interfaces ~\cite{cipcigan2016electronic,feigl2025molecular}. 

In addition to introducing and validating  the unconstrained symplectic parameterization, we benchmark the method against a Riemannian Cayley-retraction optimization framework,
highlighting the method's competitiveness in terms of computational cost, for variationally solving  Gaussian ground-states. 
In addition we examine convergence in heterogeneous systems with randomly distributed quantum Drude oscillator frequencies and polarizabilities, further highlighting the robustness of the method.


Our method for computing Gaussian ground-states is based on iteratively evaluating traces of matrix–matrix products, leading to cubic scaling in the system size per evaluation. When only the spectral gap is required, this reduces to iteratively evaluating the trace of a congruence transformation involving a tall–skinny matrix whose row dimension is independent of the system size, making the procedure structurally analogous to Krylov-type estimators~\cite{saad2011numerical}.
We observe empirical saturation for the quantum Drude lattice examples studied here, indicating empirically stable convergence behavior.
Although the method remains subject to polynomial scaling due to its reliance on matrix multiplication, its practical efficiency arises from the fact that a fixed number of matrix contractions per step is required to construct Gaussian-state correlation matrices. This structure makes the algorithm  structurally compatible with GPU acceleration \cite{rudolph2025simulating}. 

More broadly, the value of this formulation lies in providing an unconstrained, reusable Gaussian optimization primitive that can be embedded within larger or more complex computational tasks.  Accordingly, the method is not intended as a drop-in replacement for dense symplectic diagonalization in terms of wall-clock performance for single instance problems, but rather as a flexible variational component suited to iterative and composite problems, with the flexibility for future work to develop sparsity exploiting and tensor network algorithms for large and weakly non-quadratic bosonic systems.

\section{Symplectic diagonalization}
Quadratic bosonic Hamiltonians, in the absence of linear terms, can be written compactly as
\begin{equation}\label{bosonicquad}
\hat{H}=\frac{1}{2}\boldsymbol{\hat{q}}^{T}H\boldsymbol{\hat{q}} \quad{} H>0.
\end{equation}
where $H$ is the positive definite Hamiltonian matrix, and the positive definite constraint is written here as $H>0 \leftrightarrow  \boldsymbol{x}^{T}H\boldsymbol{x} >0, \forall \boldsymbol{x} \in \mathds{R}^{2d} \backslash 
\{\boldsymbol{0}\}$ and $\boldsymbol{\hat{q}}^{T}=(\hat{\chi}_{1},...,\hat{\chi}_{d},\hat{\mathcal{P}}_{1},...,\hat{\mathcal{P}}_{d})$. 

The $2d \times 2d$ matrix $\sigma_{2d}$, encodes the canonical commutation relations (CCR), 
\begin{equation}\label{sigmasymplectic}
\sigma_{2d}=\begin{bmatrix} 0 & \mathds{1}_{d} \\-\mathds{1}_{d} & 0  \end{bmatrix},
\end{equation}
 where
$[\hat{\boldsymbol{q}}_{i},\hat{\boldsymbol{q}}_{j}]=\mathrm{i}\sigma_{ij}$,  setting $\hbar=1$.
A matrix which preserves the CCR and thus $\sigma_{2d}$ under a transformation is a symplectic matrix \cite{serafini2017quantum}, belonging to the symplectic matrix group,
\begin{equation}\label{symplecticcondititon}
    S \sigma_{2d} S^{T}=\sigma_{2d} \quad{} S\in \mathcal{S}_{p}(2d,\mathds{R}).
\end{equation}
This definition can also be  generalized to rectangular symplectic matrices  \(\mathcal{S}_{p}(2k \times 2d ,\mathds{R})\), which are defined by
$S\sigma_{2d}S^{T}=\sigma_{2k},
\quad S \in  \mathcal{S}_{p}(2k \times 2d ,\mathds{R}),$
where \(\sigma_{2k}\) denotes the symplectic matrix given in Eq.~\eqref{sigmasymplectic} but restricted to a \(2k\)-dimensional subspace i.e $\sigma_{2k}=\begin{bsmallmatrix}0&\mathds{1}_k\\-\mathds{1}_k&0\end{bsmallmatrix}$, with $1\le k\le d$. Henceforth, we write $\sigma$ omitting the dimensional subscript.

Williamson's theorem \cite{williamson1936algebraic} states there is always a $2d\times 2d$ symplectic matrix,  which brings the positive definite matrix $H$ into diagonal form, $SHS^{T}=\mathrm{diag}(\epsilon_{1},\epsilon_{1},..,\epsilon_{d},\epsilon_{d})$,
where $\epsilon_{i}$ are the symplectic eigenvalues of $H$, each repeated once along the diagonal. The set of symplectic eigenvalues or symplectic spectrum is unique for a given $H$~\cite{de2006symplectic,son2021computing}.
The ground-state energy of $\hat{H}$ is given by the sum of the symplectic eigenvalues \cite{schuch2006quantum,serafini2017quantum}, 
\begin{equation}
     E_{0}=\frac{1}{2}\sum^{d}_{i=1}\epsilon_{i}=\frac{1}{4}\mathrm{tr}( SHS^{T}).
\end{equation}
The energy gap,  the energy deviation between the ground-state and the first excited state, is given in terms of the symplectic eigenvalues as, $\Delta=\mathrm{min}_{i} \epsilon_{i}$. The vanishing energy gap, where $\Delta\rightarrow 0$, is called the critical limit.
The symplectic eigenvalues and thus the ground-state energy are computable from the Hamiltonian matrix and $\sigma$ as $ \epsilon_{i}=\sqrt{\mathrm{eig}_{i}(\sigma H \sigma^{T} H)}$ ~\cite{schuch2006quantum} or alternatively $\epsilon_{i}=\mathrm{eig}_{i}(|\mathrm{i} \sigma H|)$~\cite{pirandola2009correlation}.

\section{Gaussian states}
Gaussian states are the ground and thermal states of quadratic Hamiltonians in Eq. \eqref{bosonicquad}. 
The ground-state with density matrix $\hat{\rho}$, is completely characterized by its first moments, which are here set to zero without loss of generality, and a correlation matrix (CM) of two point correlation functions \cite{schuch2006quantum,weedbrook2012gaussian}. The CM of correlation functions is given by the real, symmetric, $2d \times 2d$ matrix $\gamma$, an element of which is given in terms of the Gaussian density matrix as 
\begin{equation}\label{densitymatrixgamma}
    \gamma_{ij}=\mathrm{tr}\left[\hat{\rho}\{\boldsymbol{\hat{q}}_{i},\boldsymbol{\hat{q}}_{j}\}^{+}\right], 
\end{equation}
where $\{\cdot, \cdot\}^{+}$ is the anti-commutator \cite{schuch2006quantum}. 
 The uncertainty principle on the quadratures, can be written in matrix form, in the so-called Robertson-Schr\"{o}dinger uncertainty relation, a multimodal matrix representation of the Heisenberg uncertainty principle,
\begin{equation}\label{cmconst}
    \gamma + \mathrm{i}\sigma  \geq 0.
\end{equation}
The constraint on the CM, given in Eq. \eqref{cmconst} distinguishes it from the CM of a multivariate Gaussian probability distribution. 

Using the definition of the variational principle expressed in terms of the density matrix and Eq.~\eqref{densitymatrixgamma}, we can formulate the variational problem for a quadratic bosonic Hamiltonian as follows~\cite{schuch2006quantum}:
\begin{equation}
\begin{aligned}
E_{0} 
&= \inf_{\gamma}\; \tfrac{1}{4}\operatorname{tr}(\gamma H)
\quad \text{s.t.} \quad \gamma + \mathrm{i}\sigma \ge 0, \\
\gamma_{0} 
&= \operatorname*{arg\,inf}_{\gamma}\; \tfrac{1}{4}\operatorname{tr}(\gamma H)
\quad \text{s.t.} \quad \gamma + \mathrm{i}\sigma \ge 0,
\end{aligned}
\label{gsenergy}
\end{equation}
where the constraint $\gamma + \mathrm{i}\sigma \ge 0$ enforces the uncertainty principle, ensuring that the argument of Eq.~\eqref{gsenergy} represents a valid CM. 

\section{Gaussian Ansatz and Symplectic Optimization}
The first result of this work builds on known parameterizations of Gaussian state CMs to establish a symplectic matrix parameterization, which is ideally suited for solving the variational problem in Eq.~\eqref{gsenergy}.

\begin{figure*}
 \centering
\includegraphics{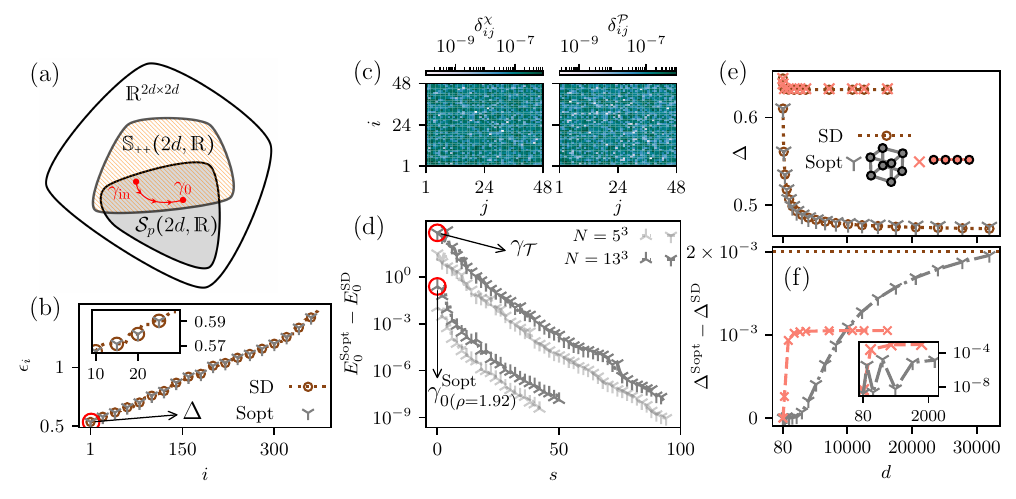}
\caption{
(a) Schematic of the Gaussian variational manifold. The ground-state CM $\gamma_{0} $ lies in the intersection of the sets of symmetric positive-definite and symplectic matrices. The symplectic optimization (Sopt) procedure performs unconstrained optimization over this manifold using unit-triangular factorization, starting from an initial CM $\gamma_{\mathrm{in}}$. 
(b--d) Validation of Gaussian ground-state recovery.
(b) Agreement between the symplectic spectrum obtained via Sopt and that from symplectic diagonalization (SD) for a $d = 3(5\times5\times5)$ cubic lattice at $\rho = 1.9$.
(c) Element-wise deviation between the CM retrieved via Sopt and that obtained from SD, shown for position correlations  (left) and momentum correlations (right), displayed as log-scale heatmaps.
(d) Convergence of the Sopt algorithm for cubic lattices of different sizes at $\rho = 1.9$, shown as the ground-state energy deviation $E^{\mathrm{Sopt}}_{0} - E^{\mathrm{SD}}_{0}$ versus the number of optimization steps $s$. Downward-pointing markers correspond to initialization from $\gamma_{\mathrm{in}}=\gamma_{\mathcal{T}}$, while upward-pointing markers  indicate a warm-started optimization initialized from the ground-state CM of the same lattice at $\rho = 1.92$. This preconditioning reduces the number of optimization steps required to reach a given accuracy by approximately a factor of two.
(e--f) Energy gap $\Delta$ obtained via Sopt for cubic and chain lattices at $\rho = 2$ and $\rho = 1.9$, respectively, compared with SD.
(f) Residual gap error after $s = 400$ optimization steps as a function of the number of lattice modes $d$. The inset highlights the low-$d$ regime.}\label{f1} 
\end{figure*}

A real symmetric matrix  is the CM of a pure Gaussian state of $d$ modes, given by $\gamma_{p}$, iff it is both symplectic and positive definite (see  Appendix \ref{A1} for proof),
\begin{equation}\label{condgammapureenrg}
    \{\gamma_{p}\} = \mathcal{S}_{p}(2d,\mathds{R}) \bigcap \mathbb{S}_{++}(2d, \mathds{R}),
\end{equation}
with $\{\gamma_{p}\} $ denoting the set of pure state CMs and
we denote the set of real symmetric positive definite matrices of dimension $2d$, as $\mathbb{S}_{++}(2d,\mathbb{R})$. We thus construct an ansatz that explicitly encodes the space of admissible CM  satisfying the uncertainty constraint. This ansatz serves as a symplectic parametrization of the feasible manifold over which the minimization in Eq.~\eqref{gsenergy} is performed, as illustrated in  Fig.~\ref{f1} (a).


To make this parametrization operational, it is necessary to specify practical methods for generating  symplectic positive definite matrices. In what follows, we present schemes based on matrix decompositions that ensure the symplectic constraint is exactly preserved. We employ factorizations in terms of unit triangular matrices, which provide a convenient and numerically stable way to parameterize elements of $\mathcal{S}_{p}(2d,\mathds{R})\bigcap \mathbb{S}_{++}(2d, \mathds{R})$ \cite{jin2020unit,jin2022optimal}.

The elementary decomposition utilized here are unit triangular matrices. The product of alternating unit triangular matrices is given as
\begin{equation}\label{unittriag}
\mathcal{L}_{\nu}= \begin{bmatrix} \mathds{1}_{d} & M_{1}\\0 & \mathds{1}_{d} \end{bmatrix}\begin{bmatrix} \mathds{1}_{d}  & 0\\M_{2} & \mathds{1}_{d} \end{bmatrix}...\begin{bmatrix} \mathds{1}_{d}  & 0\\M_{\nu-1} & \mathds{1}_{d} \end{bmatrix}\begin{bmatrix} \mathds{1}_{d}  & M_{\nu}\\0 & \mathds{1}_{d} \end{bmatrix}
\end{equation}
with $ M_{i} \in \Bbb R^{d\times d},  \: M_{i}^{T}=M_{i}.$

It has been shown in Ref. \cite{jin2022optimal} that $\{\mathcal{L}^{2}_{3} \}= \mathcal{S}_{p}(2d,\mathds{R}) \bigcap \mathbb{S}_{++}(2d,\mathds{R})$, with $\mathcal{L}^{2}_{\nu} = \mathcal{L}^{T}_{\nu}\mathcal{L}_{\nu}$.
 Given any symmetric matrix $M_{i}$ can be freely parametrized in terms of $d\times d$  upper and lower triangular matrices, or the average of an arbitrary $d\times d$ matrix and its transpose, we can freely parameterize $\gamma_{p}$, through the product $\mathcal{L}_{3}^{2}=\mathcal{L}_{3}^{T}\mathcal{L}_{3}$. This construction provides an explicit and \textit{unconstrained} parametrization of the manifold of symplectic positive definite matrices, ensuring that every choice of arbitrary matrix parameters yields a valid CM of a pure Gaussian state. Consequently, $\mathcal{L}_{3}^{2}$ serves as a convenient ansatz for generating admissible CMs that satisfy both the symplectic and positivity constraints required in Eq.~\eqref{condgammapureenrg}.
The $\mathcal{L}_{3}^{2}$ factorization uses three real symmetric  \( d \times d \) variable  matrices \( M_{1}, M_{2}, \) and \( M_{3} \) to parametrize an arbitrary pure-state CM. This requires \( 3d(d+1)/2 \) variational parameters, exceeding the \( d(d+1) \) independent parameters of a pure Gaussian-state CM~\cite{serafini2017quantum} by a factor of \( 3/2 \). Whilst this overparametrization ensures that the full manifold of pure-state CMs is covered (with a non-unique parametrization), we find that it does not adversely affect convergence in the examples considered in this article.


Using the unconstrained parameterization of the Gaussian state CM, we rewrite the variational expression for the ground-state energy in Eq.~\eqref{gsenergy} as
\begin{equation}
\begin{aligned}
E_{0} 
&= \inf_{M_{1},M_{2},M_{3}} \; \frac{1}{4}\operatorname{tr}(\mathcal{L}_{3}^{T} \mathcal{L}_{3} H),\\
(M^{0}_{1},M^{0}_{2},M^{0}_{3}) 
&= \operatorname*{arg\,inf}_{M_{1},M_{2},M_{3}} \; \frac{1}{4}\operatorname{tr}(\mathcal{L}_{3}^{T} \mathcal{L}_{3} H).
\end{aligned}\label{sympoptclasic}
\end{equation}
We have thus reformulated the Gaussian variational principle, as an unconstrained symplectic optimization (Sopt) problem, with cost function given by the trace of a series of matrix products.
The corresponding ground-state CM is then retrieved as, $\gamma_{0} = (\mathcal{L}_{3}^{0})^{T} \mathcal{L}_{3}^{0}$, where $\mathcal{L}^{0}_{3}$ is constructed from $M^{0}_{1},M^{0}_{2}$ and $M^{0}_{3}$, via Eq. \eqref{unittriag}.
The cost-function, used for computing the Gaussian ground-state in Eq. \eqref{sympoptclasic}, scales as $\mathcal{O}(d^{3})$, per optimization step $s$, where $s$ is defined as a call to the cost function. The dominant $\mathcal{O}(d^{3})$ cost per step arises from the constant number of $2d \times 2d$ sized matrix multiplications, which scale cubically with the matrix dimension. Incorporating tensor network methods and parametrized quantum circuits may further improve the scaling (see appendix \ref{B1} and \ref{C2} respectively for details)~\cite{van2025quantum}.

In addition to ground-state energy the Sopt framework can  be used to compute the spectral gap. 
This is possible via unit triangular factorization, by projecting the symplectic CM into the appropriate sub-space, thereby maintaining the rectangular symplectic constraint. We define the projection matrix $P_{1} = 
\begin{bmatrix}
1 & 0 & 0 & \cdots & 0 &  & 0 & 0 & \cdots & 0 \\[3pt]
0 & 0 & 0 & \cdots & 0 & & 1 & 0 & \cdots & 0
\end{bmatrix} \in \mathbb{R}^{2\times 2d},$
that selects the first position 
and first momentum component of $\mathcal{L}_{3}^{T} \mathcal{L}_{3}$, with $P_{1}\mathcal{L}_{3} \in  \mathcal{S}_{p}(2 \times 2d ,\mathds{R}) $. Using the cyclicality of the trace and the symplectic trace minimization theorem (see Appendix \ref{C1}), this construction yields the variational estimate,
\begin{equation}\label{SoptDELTA}
    \Delta \leq
    \inf_{m_{1},M_{2},M_{3}}
    \frac{1}{2}\mathrm{tr}\!\left(
    P_{1}\mathcal{L}_{3}H\mathcal{L}^{T}_{3}P^{T}_{1}
    \right),
\end{equation}
which we empirically show is tight for the examples benchmarked here.
While $M_{1}\in\mathbb{R}^{d\times d}$ in the full parameterization, the action of $P_{1}$ leaves only a single row $m_{1}=P_{1}M_{1}\in\mathbb{R}^{d}$ variational in Eq.~\eqref{SoptDELTA}; accordingly, the infimum is taken over $(m_{1},M_{2},M_{3})$ rather than over the full matrix $M_{1}$.

Computing the cost function in Eq. \eqref{SoptDELTA},
is dominated by multiplication between a tall-and-skinny matrix and a dense square matrix, owing to the dimensional reduction introduced by the projector \(P_{1}\). Consequently, the computational scaling per optimization step is \(\mathcal{O}(d^{2})\), which is a polynomial speedup, compared with Eq. \eqref{sympoptclasic}. 

\section{Numerical Examples}
\subsection{Crystal lattices}
To demonstrate the capabilities of Sopt for a model with a well-established solution, we consider assemblies of dipole-coupled quantum Drude oscillators (QDOs) on lattices~\cite{willby2025quantum,cao1992many,donchev2006many}. Each QDO contributes three oscillator modes, so the total number of
modes is $d=3N$ for $N$ QDOs. Each QDO is characterized by a frequency $\omega_{\mu}$ and
polarizability $\alpha_{\mu}$. The off-diagonal $3\times3$ blocks of the
potential matrix describe the interaction between QDOs $\mu$ and $\xi$,
and are given by
$
V_{\mu\xi}
=
\frac{\omega_{\mu}\omega_{\xi}}{\omega_0^2}
\sqrt{\alpha_{\mu}\alpha_{\xi}}\,
\mathcal{T}_{\mu\xi},
~ \mu\neq\xi,
$
with diagonal blocks
$V_{\mu\mu}
=\frac{\omega_\mu^2}{\omega_0^2}\mathds{1}_3$. Here $\omega_0$ is a
reference frequency. The off-diagonal blocks of the dipole--dipole interaction matrix are
$
\mathcal{T}_{\mu\xi}
=
\frac{1}{R_{\mu\xi}^3}
\left(
\mathds{1}_3
-
\frac{3\,\mathbf{R}_{\mu\xi}\otimes\mathbf{R}_{\mu\xi}}
{R_{\mu\xi}^2}
\right),
~ \mu\neq\xi,
$
with diagonal blocks $\mathcal{T}_{\mu\mu}=0_{3\times3}$, where $0_{3\times3}$ denotes
the $3\times3$ zero matrix. The resulting Hamiltonian matrix is
$H=V\oplus\mathds{1}_d$.
We first consider identical QDOs
($\omega_\mu=\omega_0$, $\alpha_\mu=\alpha$) on lattices with fixed
geometry. Introducing the dimensionless nearest-neighbour spacing
$\rho=R/\alpha^{1/3}$, where
$R=\min_{\mu\neq\xi}R_{\mu\xi}$
\cite{donchev2006many,willby2025quantum}, the interaction tensor may be
written as
$\alpha\mathcal{T}=\rho^{-3}\tilde{\mathcal{T}}$, where
$\tilde{\mathcal{T}}$ depends only on the lattice geometry.

For block–diagonal quadratic Hamiltonians with no intermodal momentum couplings, such as the QDO Hamiltonian, the ground–state CM follows directly from diagonalizing \(V\), yielding \(\gamma_{0} = V^{-1/2} \oplus V^{1/2}\) ~\cite{audenaert2002entanglement,schuch2006quantum,cramer2006entanglement}, with \(V^{\pm 1/2} = O\,\mathrm{diag}(\sqrt{\lambda^{\pm 1}_{1}},\ldots,\sqrt{\lambda^{\pm 1}_{d}})\,O^{T}\), where \(O\) diagonalizes \(V\) and \(\lambda_{i}\) are its eigenvalues. For such structured Hamiltonians, the symplectic eigenvalues are given by $\epsilon_{i}=\sqrt{\lambda_{i}}$. We additionally benchmark the method on quadratic Hamiltonians with explicit position–momentum coupling, demonstrating comparable accuracy, reported in Appendix \ref{D1}.

For a proof-of-principle demonstration of Sopt for solving QDO ensembles,
we employ a conjugate-gradient (CG) method
\cite{nocedal2006numerical} (SciPy implementation) to solve the
expression in Eq.~\eqref{sympoptclasic}. The optimization is carried out
over the independent entries of the symmetric matrices
$M_1$, $M_2$, and $M_3$, with gradients defined using the Frobenius inner
product on the Euclidean parameter space. The corresponding
gradient expressions, used to construct the Jacobian, are derived
in Appendix~\ref{E1}. In contrast to optimization methods that compute Riemannian gradients, the metric here is
introduced on the unconstrained parameterization rather than on the
symplectic manifold.

To further illustrate the straightforward trainability of the objective function, we also apply simple gradient-descent–based optimization schemes. Unless stated otherwise, we initialize the optimization with $M_1=\rho^{-3}\mathcal{\tilde{T}}$ and $M_2=M_3=0$, with resultant initialized CM $\gamma_{\mathrm{in}}=\gamma_{\mathcal{T}}$, where
$\gamma_{\mathcal{T}}=\begin{bsmallmatrix}
\mathds{1}_{d}+\rho^{-6}\mathcal{\tilde{T}}^{2} & \rho^{-3}\mathcal{\tilde{T}} \\ \rho^{-3} \mathcal{\tilde{T}} & \mathds{1}_{d}
\end{bsmallmatrix}$.  This physically motivated choice significantly accelerates convergence, acting as an effective preconditioner. Although uninformed initializations converge to the same optimum, they typically require substantially more optimization steps to do so. This does not restrict generality, since an analogous initialization can always be constructed from the interaction matrix elements of a general quadratic Hamiltonian.


We refer to the ground-state CM, energy and gap computed via symplectic optimization as \( \gamma^{\mathrm{Sopt}}_{0} \), \( E^{\mathrm{Sopt}}_{0} \) and \( \Delta^{\mathrm{Sopt}} \), distinguishing them from the ground-state CM, energy and gap obtained via symplectic diagonalization (SD). All lattice calculations are performed with open boundary conditions.
For the \( N=3\times3\times3 \) cubic lattice at \( \rho=1.9 \), the Frobenius-norm error between \( \gamma^{\mathrm{Sopt}}_{0} \) and \( \gamma^{\mathrm{SD}}_{0} \) is \( 2.4\times10^{-4} \) at CG solver tolerance \( 10^{-5} \) and \( 1.7\times10^{-6} \) at tolerance \( 10^{-7} \). At \( \rho=2.5 \), the corresponding errors are \( 4.9\times10^{-4} \) and \( 4.2\times10^{-6} \), respectively, indicating that the coupling strength does not significantly affect the accuracy.
The unique symplectic spectrum of \( H \) is given by the eigenvalues of the optimized symplectic form of the Hamiltonian, \( \mathcal{L}^{0}_{3} H (\mathcal{L}^{0}_{3})^{T} \), and is shown in Fig.~\ref{f1}(b) to overlap with the spectrum computed by SD at tolerance \( 10^{-5} \), with a Frobenius-norm error between the two diagonal matrices of \( 3.0\times10^{-5} \). This demonstrates that \( \gamma^{\mathrm{Sopt}}_{0} \) contains an accurate encoding of the symplectic spectrum of $H$. Fig.~\ref{f1}(c) shows the element-wise agreement between
the position CM block given by $\delta^{\chi}_{ij} = \lvert (\Pi_{1}^{T}\gamma^{\mathrm{Sopt}}_{0}\Pi_{1})_{ij} - (V^{-1/2})_{ij} \rvert$ and for the momentum block $\delta ^{\mathcal{P}}_{ij} = \lvert (\Pi_{2}^{T}\gamma^{\mathrm{Sopt}}_{0}\Pi_{2})_{ij} - (V^{1/2})_{ij} \rvert$, where $\Pi^{T}_{1}=[\mathds{1}_{d},0_{d}]$ and $\Pi^{T}_{2}=[0_{d},\mathds{1}_{d}]$ , computed at tolerance $10^{-7}$, and the ground-state CM obtained via diagonalization for a square lattice. The maximum errors in the two-point correlation functions are $\max(\delta^{\chi}_{ij}) = 8.4\times10^{-7}$ and $\max(\delta^{\mathcal{P}}_{ij}) = 8.2\times10^{-7}$, both on the order of the solver tolerance.

In Fig.~\ref{f1}(d), we show the convergence of Sopt on cubic lattices of different sizes and with different initializations. Initializing the optimization with \( \gamma_{\mathcal{T}} \), we find that the number of steps required to reach comparable accuracies given by \( E^{\mathrm{Sopt}}_{0}-E^{\mathrm{SD}}_{0}\) increases only modestly as the cubic lattice size grows. After 30 steps, the error for the \( N = 5 \times 5 \times 5 \) cubic lattice is \( 1\times10^{-3} \), while for the order-of-magnitude larger \( N = 13 \times 13 \times 13 \) instance, a lower error of \( 9.8\times10^{-4} \) is achieved after only 12 additional steps. 
Whilst not shown, we also consider \( N=10\times10 \) and \( N=40\times40 \) square lattices, where after 108 and 112 steps respectively the error in the ground-state energy is \(2.1\times10^{-9} \) and \( 6.6\times10^{-8} \). For both the square and cubic lattice, and for all considered sizes, 65 steps are sufficient to reduce the deviation from diagonalization below \( 10^{-4} \). Fig.~\ref{f1}(d) further shows the convergence on the cubic lattice with a warm start. The initialization for the Sopt minimization is given by the ground-state CM of the same lattice setup at a \( 1\% \) higher \( \rho \) value. This preconditioning reduces the number of steps required to reach comparable accuracy to the \( \gamma_{\mathcal{T}} \) initialization by approximately a factor of two, where 44 and 51 steps are required to reach accuracies of \( 1.4\times10^{-8} \) and \( 2\times10^{-9} \) for the \( N = 5 \times 5 \times 5 \) and \( N = 13 \times 13 \times 13 \) cases, respectively.


In Fig.~\ref{f1} (e)–(f) we employ Sopt to determine the energy gap via Eq.~\eqref{SoptDELTA}. While the chosen symplectic parameterization ensures that the global minimum of the variational objective coincides with the exact ground-state energy, no analogous analytic guarantee exists for extracting the spectral gap with unit triangular parametrization; nevertheless, we observe numerical convergence across all lattice sizes studied. We fix the number of optimization steps, $s = 400$, and consider a chain and cubic lattices. The Sopt algorithm is applied to determine the energy gap $\Delta$ in each case, using gradient descent with momentum (learning rate $0.26$, momentum parameter $0.95$). Fig.~\ref{f1} (e) shows the convergent energy gap in the chain and the more gradually converging energy gap in the cubic lattice with increasing number of lattice modes. 
The lattice size dependent behavior in the energy gap is reflected in the convergence of Sopt, where the residual error,  $\Delta^{\mathrm{Sopt}}-\Delta^{\mathrm{SD}}$, is shown in Fig.~\ref{f1} (f).  For the one-dimensional chain, we observe rapid saturation as a function of system size, with the error in the Sopt solution remaining constant to three decimal places at $1\times10^{-3}$, for $d\geq 900$.  For the cubic lattice, the error increases beyond that of the chain, as a function of lattice size, however it remains below $2\times 10^{-3}$ at $d =31,944$ modes. 
\subsection{Sweeps and method comparison}\label{sweepsec}
\begin{figure}
    \centering
    \includegraphics{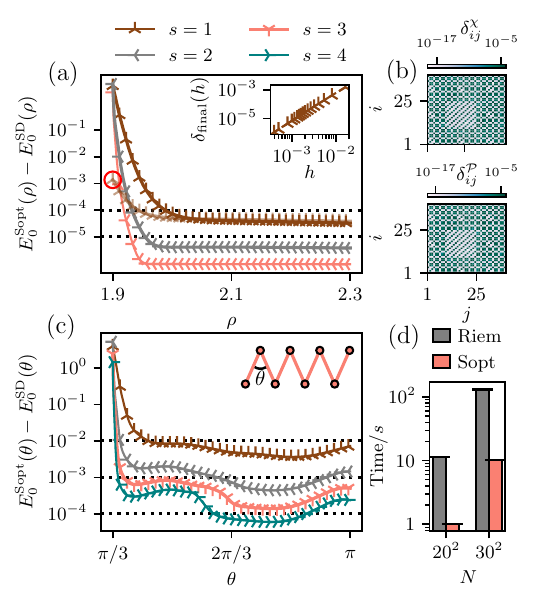}
    \caption{Warm-started parameter sweeps and method comparison. (a) Uniform $\rho$-sweep on a $10\times10$ square lattice over $\rho\in[1.9,2.3]$, showing the ground-state energy error after a fixed number of optimization steps per $\rho$ point. Each optimization is initialized using the  variational parameters, computed from the preceding value of $\rho$. The inset shows the energy error attained at the final $\rho$ point in the sweep, as a function of sweep-spacing,  $h$.
    The red marker indicates the initial point, which is optimized for 30 iterations before subsequent points are propagated using only a single optimization step per sweep point. (b) Element-wise deviation between the CM reconstructed after the sweep with three optimization steps per $\rho$ point and the CM obtained via SD, shown for the position (top) and momentum (bottom) correlation blocks. (c) Geometric sweep parameterized by the angle $\theta$ over $\theta \in [\pi/3,\pi]$, showing the ground-state energy error for different numbers of steps per geometry point as the lattice geometry is varied from the zigzag to linear configuration, at $\rho=1.9$, for a 100 QDO chain. (d) Mean runtime over 5 repeated runs of a single optimization step for the Riemannian Cayley-retraction method of Ref.~\cite{son2021computing} (labelled Riem here) and Sopt on different sized square lattices, recorded on a dual-socket AMD EPYC 7763 (128-core) compute node using a single CPU core.}
    \label{sweeps}
\end{figure}

To assess the effectiveness of warm-starting, we consider parameter sweeps in which previously converged solutions are reused following small changes in a Hamiltonian parameter. In Fig.~\ref{sweeps}(a)  we fix the number of optimization steps and apply the Sopt algorithm, using gradient descent with momentum (learning rate 0.1, momentum parameter 0.8) at each point in the sweep. We consider a sweep over uniformly spaced dimensionless nearest-neighbor spacings
$\{\rho_j\}_{j=1}^{n_\rho}$. After a fixed number of optimization steps at $\rho_j$, the optimized
variational parameters are used to initialize the optimization at the
subsequent sweep point, $\rho_{j+1}$, with sweep spacing given by $h=\frac{\rho_{n_{\rho}}-\rho_{1}}{n_{\rho}-1}$.

Fig.~\ref{sweeps}(a) demonstrates that warm-started optimization accurately
tracks the ground-state across the $\rho$-sweep. When only a
single gradient descent step is applied at every value of \(\rho_{j}\), \(77.8\%\) of the sweep points achieve
ground-state energy errors below \(10^{-4}\).
When the sweep is instead initialized by performing \(30\) gradient
descent steps at the first point, $\rho_{1}$, and then only a single
step is used at each subsequent value, \(\rho_{j>1}\), this fraction increases to
\(89.2\%\). Allowing 2 gradient descent steps at
every $\rho_{j}$ further improves the accuracy, in comparison to the single step instance, with \(86.2\%\) of sweep
points achieving errors below \(10^{-5}\). Notably, this accuracy is maintained
across a $\rho$ interval for which the correlation energy changes by
approximately \(70\%\). The inset of Fig.~\ref{sweeps}(a) shows the final energy error $\delta_{\mathrm{final}}=E^{\mathrm{Sopt}}_{0}(\rho_{n_{\rho}})-E^{\mathrm{SD}}_{0}(\rho_{n_{\rho}})$ as a function of sweep spacing $h$. The sweep resolutions used to generate the inset are listed in
Appendix \ref{F1}. The final error scales as $
\delta_{\rm final}\propto (h)^{1.96}
$ (\(R^2=0.9998\)), consistent with quadratic scaling in the sweep spacing.

Fig.~\ref{sweeps}(b) shows accurate reconstruction of the ground-state CM from the Sopt sweep, with the element-wise error in ground-state CM at the final $\rho_{n_{\rho}}$ value in the 3-step sweep, shown via the comparison with SD, with $\delta^{\chi}_{ij},\delta^{\mathcal{P}}_{ij} <9\times10^{-5}~ \forall i,j \in d$.
In addition to density sweeps, we consider a geometric sweep parameterized by an angle $\theta$, which continuously deforms the system from a zigzag configuration ($\theta=\pi/3$) to a  linear chain configuration ($\theta=\pi$). In Fig.~\ref{sweeps}(c) we consider 500 uniformly spaced points, between  the zigzag and linear configuration. The zigzag configuration is illustrated in the inset of Fig.~\ref{sweeps}(c). Using  4 optimization steps per geometry, $97.6\%$ of sweep points attain energy errors below $10^{-3}$.

While warm-starting reduces the number of optimization steps required across a parameter sweep, the practical efficiency of the approach also depends on the cost of each step.
To place the computational cost of Sopt in context, we compare against the Cayley-transform method of Ref.~\cite{son2021computing}, which was found in Ref.~\cite{gao2021riemannian} to outperform the quasi-geodesic solver for the symplectic eigenvalue problem. This provides a natural optimization-based baseline for Sopt, since both approaches solve the same underlying symplectic trace-minimization problem. For both methods, the cost of a single optimization step scales as $\mathcal{O}(d^{3})$, due to dense matrix operations. However, the Cayley-transform update is performed directly on a dense
$2d\times 2d$ symplectic matrix and requires the application of an
inverse linear operator, implemented here as a dense linear solve with
multiple right-hand sides, arising from the Cayley retraction (see Appendix \ref{G1} for further information and numerical details). As shown in Fig.~\ref{sweeps}(d), this results in the Riemannian Cayley-retraction method having an order of magnitude \textit{higher} wall-clock cost per optimization step, in comparison with Sopt, for both lattice sizes considered. 

\subsection{Heterogeneous parameters}
\begin{figure}
    \centering
\includegraphics{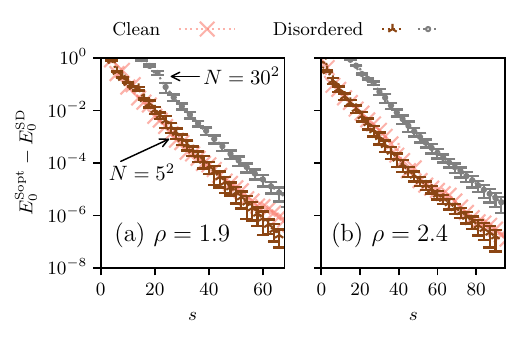}
    \caption{Comparison of convergence in clean and disordered lattices at different $\rho$ values on a $5\times5$ square lattice. The grey open circular markers additionally show the mean convergence for a disordered $30\times30$ square lattice averaged  to assess convergence at larger scales. Disorder is introduced through independently sampled oscillator frequencies and polarizabilities, where $\omega_{\mu}\sim\mathcal{U}(\omega_{\mathrm{min}},\omega_{\mathrm{max}})$ and $\alpha_{\mu}\sim\mathcal{U}(\alpha_{\mathrm{min}},\alpha_{\mathrm{max}})$, with $\alpha_{\mu}\sim\mathcal{U}(0.95,1)$ and $\omega_{\mu}\sim\mathcal{U}(1,1.05)$ in $\mathrm{(a)}$ and $\alpha_{\mu}\sim\mathcal{U}(1,2)$ with fixed frequency in $\mathrm{(b)}$. The disordered results correspond to averages over 50 disorder realizations for the $5\times5$ lattice and 100 disorder realizations for the $30\times30$ lattice, with error bars indicating one standard deviation, plotted on a log scale. Despite substantial parameter heterogeneity, the convergence trajectories remain nearly indistinguishable from those of the corresponding clean lattices.}
    \label{disorder}
\end{figure}

To investigate the effect of parameter heterogeneity on the optimization, we introduce disorder in the QDO frequencies and polarizabilities by independently sampling $\omega_\mu$ and $\alpha_\mu$ from uniform distributions, the results of which are shown in Fig. \ref{disorder}. For the disordered systems, the lattice geometry is held fixed at that of the corresponding clean lattice, with $\rho$ referring to the dimensionless spacing of that clean configuration
The initialization is modified accordingly by replacing
$M_1=\rho^{-3}\tilde{\mathcal{T}}$ with the heterogeneous interaction
matrix, while keeping $M_2=M_3=0$.
For the $5\times5$ lattice, the mean convergence of the disordered realizations remains close to that of the corresponding clean system, even for substantial disorder amplitudes, including a two-fold variation in polarizability in panel (b).
The error bars indicate that individual disorder realizations exhibit only modest variability and follow the same overall convergence trend as the clean system.
To assess the robustness of this behavior at larger system sizes, we additionally consider a $30\times30$ lattice averaged over 100 independent disorder realizations. The corresponding error bars indicate only modest variability between realizations throughout the optimization. For the disorder distributions considered here, the standard deviation decreases from $\mathcal{O}(1)$ during the initial iterations to $\mathcal{O}(10^{-5})$ nearer to convergence.



\section{Conclusion} 
This work establishes a constraint-free reformulation of Gaussian ground-state optimization. In particular we have shown how the Gaussian variational problem can be recast as a symplectic optimization problem over the space of positive-definite symplectic matrices. We then numerically demonstrated how to use symplectic optimization to find Gaussian ground-states via unit triangular factorization of positive-definite symplectic matrices. This results in a globally unconstrained symplectic trace minimization problem, which we carried out numerically for systems of dipole-coupled quantum Drudes on lattices.

Our symplectic optimization approach retrieves ground-state energies, correlation matrices, and spectral gaps in excellent agreement with symplectic diagonalization. While the benchmarks presented here correspond to quadratic Hamiltonians whose symplectic spectra are expressible as simple functions of the normal eigenvalues of the coupling matrix, the framework itself is fully general and directly applicable to non–block-diagonal quadratic Hamiltonians.

Beyond single-instance optimization, we demonstrated that warm-started Sopt accurately tracks ground-states across both relative density and geometry sweeps using only a small number of optimization updates per parameter value. We further benchmarked the method against a Riemannian Cayley-retraction optimizer and examined its behavior in heterogeneous systems with substantial variations in oscillator frequencies and polarizabilities. Across all cases considered, the convergence behavior remained robust and the recovered Gaussian states remained in excellent agreement with symplectic diagonalization.

Whilst we have here focused on model quantum Drude systems, the methods can straightforwardly be extended to realistic materials, within the many-body dispersion framework. While dense symplectic diagonalization remains highly effective for isolated quadratic Hamiltonians, the present framework is particularly attractive in settings involving large families of related configurations, where previously optimized solutions can be reused. Understanding the extent to which this solution reuse can provide practical advantages in realistic many-body dispersion calculations remains an important direction for future work. Preliminary results presented in Appendix \ref{H} further indicate that warm-started Sopt remains effective under aggressive banded parameter truncation, suggesting a practical route towards reducing both the computational cost and memory requirements of the optimization.
Such extensions could be of particular use, in exploring many-body quantum effects outside the realm of fixed lattice positions.
For instance going beyond pairwise Stockmayer fluid descriptions \cite{groh1996density,varghese2025dynamic} via incorporating many-body quantum effects, across multiple configurations, would enable more accurate modeling of fluid-mediated self-assembly ~\cite{serwatka2023quantum,singh2025self}.

Extensions of the method could explore approximating perturbative non-quadratic terms \cite{shi2018variational,guaita2019gaussian} via symplectic optimization, utilizing the matrix parameterizations developed here. Such approaches could be useful for including higher-order multipole interactions between quantum Drudes at scale. A natural next step is to impose parameter reductions in the unit-triangular construction. Many physically relevant quadratic Hamiltonians are sparse, banded, or low-rank. Provided accuracy is retained under aggressive parameter reduction, dominant cost and gradient contractions could be reduced to sparse–sparse matrix multiplications. Such structure may also admit efficient tensor-network formulations and corresponding variational quantum algorithm implementations \cite{lubasch2020variational,siegl2026tensor}, the full development of which we leave for future work.
Finally, the proposed correlation-matrix parameterization may find applications in foundational studies, optimizing over  the Gaussian-manifold, for problems with no known closed form solution ~\cite{camargo2021entanglement,aurell2024random,walschaers2021non}.

\section*{Acknowledgements}
Simulations were run
on the University of Oxford Advanced Research Computing (ARC) facility. 
This work has received funding from the European Union’s Horizon Europe research and innovation program (HORIZON-CL4-2021DIGITAL-EMERGING-02-10) under grant agreement No. 101080085 QCFD.
DJ and TH are partly funded by the Cluster of Excellence ‘Advanced Imaging of Matter’ of the Deutsche Forschungsgemeinschaft (DFG)—EXC 2056- project ID390715994. 
DJ acknowledges the support by DFG project “Quantencomputing mit neutralen Atomen” (JA 1793/1-1, JapanJST-DFG-ASPIRE 2024) and the Hamburg Quantum Computing Initiative (HQIC) project EFRE. The EFRE project is co-financed by ERDF of the European Union and by “Fonds of the Hamburg Ministry of Science, Research, Equalities and Districts (BWFGB)”.
\newline

\appendix


\section{Pure State CM Parameterization}\label{A1}

The positivity constraint on the density matrix is enforced in phase-space by the  Robertson-Schr\"{o}dinger uncertainty relation on the CM. 
For a \textit{pure} Gaussian state, with CM given by $\gamma_{p}$, we have the following extra condition on the CM \cite{wolf2004gaussian,hiroshima2007monogamy},
\begin{equation}\label{ispuredet}
    \mathrm{Det}(\gamma_{p})=1 \leftrightarrow \left(\gamma_{p} \sigma \right)^{2} = -\mathds{1}_{2d}.
\end{equation}
This condition provides a general parameterization for pure state CMs:


\begin{lemma}\label{lem1}
A real symmetric matrix $\gamma_{p}$ is the covariance matrix of a pure Gaussian state of $d$ modes iff there exist real symmetric $d \times d$ matrices $X$ and $Y$ with $X > 0$ such that
\begin{equation}\label{purecm}
    \gamma_{p}=\begin{bmatrix}
        X & XY \\ YX & YXY+X^{-1}
    \end{bmatrix},\quad{} X>0, \:Y=Y^{T}   \:\& \: X=X^{T}.
\end{equation}
\end{lemma}

See~\cite{wolf2004gaussian} for proof.


To prove the pure CM parametrization, utilized in the main-text, we make use of the following;

\begin{lemma}\label{lem2}
Any symmetric positive definite symplectic matrix can be written as
\begin{equation}\label{symplecticposdef}
\mathcal{S}_{p}(2d,\mathds{R}) \bigcap \mathcal{S}_{++}(2d, \mathds{R})
=\left\{ \begin{bmatrix}
    \mathds{1}_{d} & 0 \\
    Y & \mathds{1}_{d}
\end{bmatrix}
\begin{bmatrix}
    X & 0 \\
    0 & X^{-1} 
\end{bmatrix}
\begin{bmatrix}
    \mathds{1}_{d} & Y \\
    0 & \mathds{1}_{d}
\end{bmatrix} \right\},
\end{equation}
where $\mathcal{S}_{++}(2d, \mathds{R})$ denotes the set of matrices which are both positive definite and symmetric.  
Here $X$ is symmetric ($X = X^{T}$) and positive definite ($X > 0$), and $Y$ is symmetric.
\end{lemma}

See~\cite{dopico2009parametrization} for proof.

By expanding out the expression in Eq. \eqref{symplecticposdef} and comparing to the parametrization for pure Gaussian state CMs in Eq. \eqref{purecm}, we prove the result stated in the main-text, that a symmetric positive definite symplectic matrix always constitutes a valid pure state CM. 
\section{Tensor Network Extension to the Method}\label{B1}

\begin{figure*}[t]
   \centering
   \includegraphics[scale=1.0]{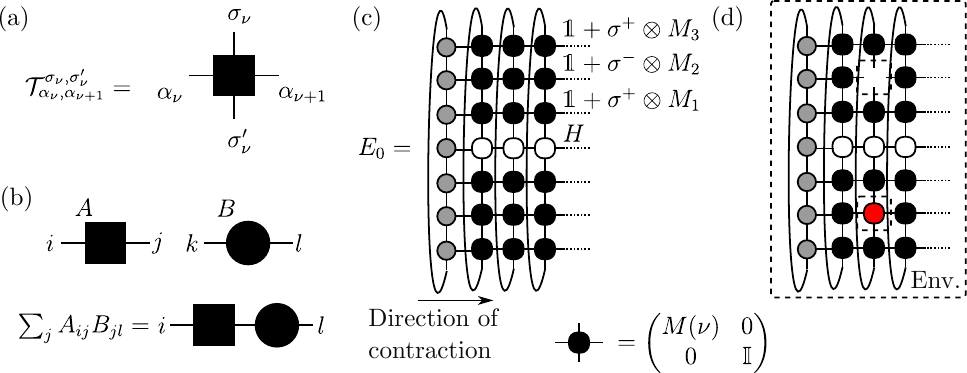}
   \caption{(a) The graphical representation of a tensor $T^{\sigma_\nu,\sigma_\nu'}_{\alpha_{\nu},\alpha_{\nu+1}}$.
   (b) The graphical representation of the product of two matrices $A$ and $B$. 
   The connection of two legs represents the contraction of shared matrix dimensions.
   (c) Tensor network diagram of the cost function $E_0$. 
   (d) networks required to compute the gradient of $E_0$ with respect to a local tensor, 
   in this case, $M_2(2)$. The region wrapped by the dotted line containing two holes (one with a red tensor)
   is the environment of the local tensor $M_2(2)$. 
   The full gradient is computed by summing term shown
   where the red tensor is replaced by $\begin{pmatrix} M_2(2)  & 0\\0 & 0 \end{pmatrix}$, 
   the term where the red tensor is replaced by $\begin{pmatrix} 0 & 0\\0 & \mathbb{I} \end{pmatrix}$, 
   and their transposes. 
   Here, $\mathds{1}$ is an identity matrix of the appropriate dimension
   and $\mathbb{I}$ is the $2\times 2$ identity matrix.}\label{fig:tensroentwork}
\end{figure*}

In this appendix, we argue that the computational intractability of a large number of modes on current hardware 
can be addressed by decomposing the matrices $M_\nu$ and $H$ into chains of tensors, called tensor networks (TN)
\cite{10.5555/2011832.2011833,pirvuMatrixProductOperator2010,schollwockDensitymatrixRenormalizationGroup2011},
and optimizing individual tensors.
To achieve this, we first discuss how a general $d\times d$ matrix $\mathcal{M}$ 
is decomposed into contractions of tensors of rank 4, $\mathcal{T}^{\sigma_n,\sigma_n'}_{\alpha_n,\alpha_{n+1}}$. 
To this end, we find $\log_2 d$ tensors of the above form that give the element $i,j$ of $\mathcal{M}$ as 
\begin{align}
   \mathcal{M}[i,j] &= \sum_{\alpha_1,\alpha_2,\cdots\alpha_{\log_2 d-1}} 
   \mathcal{T}^{\sigma_1,\sigma_1'}_{\alpha_1,\alpha_{2}}\cdots
   \mathcal{T}^{\sigma_1,\sigma_2'}_{\alpha_1,\alpha_{3}}
   \mathcal{T}^{\sigma_n,\sigma_n'}_{\alpha_n,\alpha_{n+1}}
\end{align}
where $\sigma_n$ and $\sigma_n'$ correspond to the $n$\textsuperscript{th} bits of the binary representation of the 
indices $i$ and $j$ respectively ($i,j\in\{0,1,\cdots \log_2d-1\}$). 
We see that the size of $\mathcal{T}^{\sigma_n,\sigma_n'}_{\alpha_n,\alpha_{n+1}}$ is at worst, 
$2\times2\times\chi^2$, where the $\chi$ is the bond dimension (the largest range spanned by the bond indices $\alpha_1,\cdots,\alpha_{\log_2d-1}$).
By choosing a reasonable value for $\chi$, the number of variational parameters to be optimized is reduced to
$\mathcal{O}(\chi^2\log_2 d)$, offering a potential exponential speed up for the ground state search.

To create $\mathcal{L_\nu}$ from Eq. \eqref{unittriag} in the main-text,
we observe that (upper/lower) triangular matrices in the product are identical to
$(\mathcal{I} + \sigma^{(+/-)} \otimes M_\mu)$, where
\begin{align}
   \sigma^+ = \begin{pmatrix}
      0 & 1 \\
      0 & 0
   \end{pmatrix},
   \sigma^- = \begin{pmatrix}
      0 & 0 \\
      1 & 0
   \end{pmatrix}. 
\end{align}
In terms of tensor networks, this is equivalent to prepending the tensor of bond dimension $1$ that contains the appropriate $\sigma^{\pm}$, 
and placing the identity matrix in the lower right corner of each tensor after expanding the bond dimension of each tensor by $1$.

In Fig.~\ref{fig:tensroentwork}(a-d), we show the graphical notation for tensors and their contractions. 
In this notation, a tensor and its indices are denoted by a square and its legs, where the legs represent the indices. 
The contraction of two tensors is represented by two legs being connected (Fig.~\ref{fig:tensroentwork}a). 
As an example, in Fig.~\ref{fig:tensroentwork}b, we show the graphical representation of a product of two matrices.
With this representation, the ground state energy, $E_0$, (main-text Eq. \eqref{sympoptclasic}) can be represented as it is shown in Fig.~\ref{fig:tensroentwork}c. 
The derivative of $E_0$ with respect to the $\nu$\textsuperscript{th} local tensor 
$M_1(\nu)=\mathcal{T}^{\sigma_\nu,\sigma_\nu'}_{\alpha_\nu,\alpha_{\nu+1}}$ of $M_1$ 
(similarly for $M_2$ and $M_3$) then becomes a contraction of all the other local tensors except for $M(\nu)$ 
as shown in Fig.~\ref{fig:tensroentwork}d.

Based on this, we find that the complexity for evaluating $E_0$ is $\mathcal{O}(\chi^7\log_2 d)$. 
This optimal complexity is achieved when the contraction is performed in the direction indicated by the arrows
in Fig.~\ref{fig:tensroentwork}c. 
Furthermore, to compute the derivative with respect to a local tensor, 
$\mathcal{O}(\chi^8)$ floating-point multiplications are required to create the environment (Fig.~\ref{fig:tensroentwork}d, contracted network surrounded by the dotted line), 
and an additional $\mathcal{O}(\chi^4)$ is needed to compute the derivative.

\begin{figure}[t]
   \centering
\includegraphics{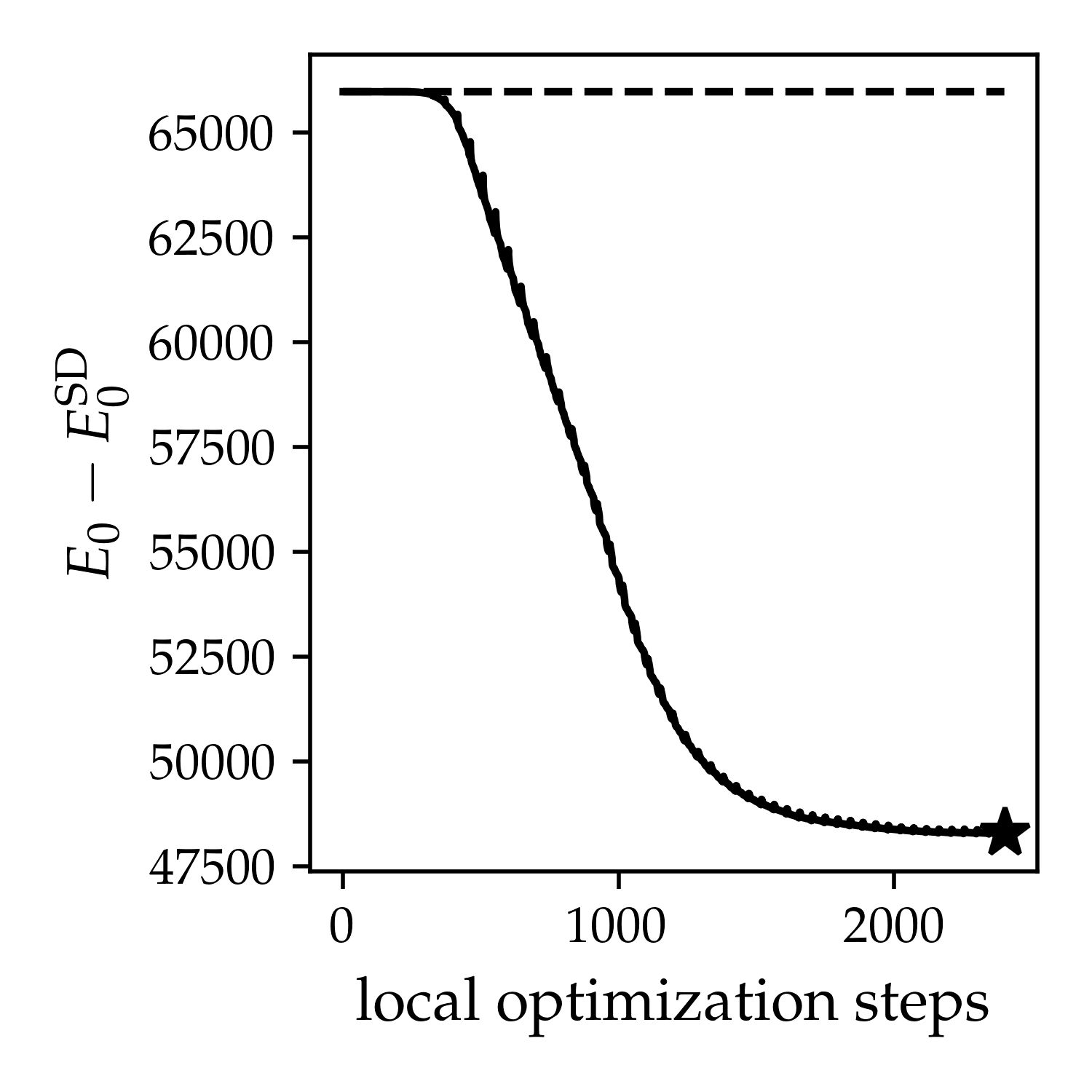}
   \caption{
      The ground state energy of the bosons on a $d=256\times256\times256$ cubic lattice, 
      computed with the optimization of local tensors with $\chi=4$ using the L-BFGS 
      algorithm (10 modes, with 10 iterations) \cite{zhuAlgorithm778LBFGSB1997}. 
      We set the potential energies to unity and assigned a weight of $0.1$ to the nearest-neighbor interaction.
      The dashed line represents the trivial bound on the ground state energy, $d/2$. 
      The optimization converged to within $\approx 0.8\%$ of the exact ground state energy 
      $E^{\mathrm{SD}}_0\approx 8.3226\times10^{6}$. 
}\label{fig:tnoptresult}
\end{figure}

In Fig.~\ref{fig:tnoptresult}, we show the optimization results for quantum harmonic oscillators, restricted to fluctuate in one spatial dimension, residing on a cubic lattice of 
dimensions $d=256\times256\times256$.
We set the potential energies to unity and assigned a weight of $0.1$ to the nearest-neighbor interactions.
As shown, we found a configuration with an energy within $\approx 0.8\%$ of the exact energy, which is 
well below the trivial bound of $d/2$ (dashed line). 
The computation of $24\times100$ local optimization steps indicated by the star, took $\approx 15$ 
hours on a single CPU (Intel(R) Xeon(R) CPU E5-2698 v4 @ 2.20GHz). 

The true ground state energy is computed analytically by finding the exact 
expressions for the eigenvalues of the $256 \times 256$ matrix $J$,
where the first off-diagonal elements are set to $0.1$ to model a 1D nearest-neighbor chain.
The $k$-th eigenvalue of $J$ is expressed as $0.2 \cos\frac{k\pi}{\sqrt[3]{d}+1}$ \cite{yueh2005eigenvalues}.
Since the Hamiltonian is structured as a Kronecker sum of these matrices,
\begin{equation}
   H = I + J \otimes I \otimes I + I \otimes J \otimes I + I \otimes I \otimes J
\end{equation}
its eigenvalues are determined by all possible triplets of the eigenvalues of $J$.
The final energy is obtained by summing the square roots of these combined eigenvalues across its entire 
spectrum.
\section{Bloch--Messiah parameterization for quantum optimization}\label{C2}
The Bloch--Messiah decomposition expresses any pure-state CM as~\cite{serafini2017quantum}
\begin{equation}
\label{gamdecom}
\gamma_p=O_s\tilde{D}O_s^T,
\end{equation}
where
\[
O_s\in\mathcal{S}p(2d,\mathds{R})\cap SO(2d,\mathds{R}),
\]
and
\[
\tilde D
=D\oplus D^{-1},
\qquad
D=\mathrm{diag}(d_1,\ldots,d_d),
\qquad
d_i>0.
\]
The orthogonal symplectic matrix $O_s$ may be parameterized by an arbitrary unitary matrix $U$ as~\cite{serafini2017quantum}
\begin{equation}
\label{uandustar}
O_s=
\bar U^\dagger
\begin{pmatrix}
U^\star&0\\
0&U
\end{pmatrix}
\bar U,
\qquad
\bar U=
\frac{1}{\sqrt2}
\begin{pmatrix}
\mathds1&i\mathds1\\
\mathds1&-i\mathds1
\end{pmatrix}.
\end{equation}
Substituting Eq.~\eqref{uandustar} into Eq.~\eqref{gamdecom} gives
\begin{equation}
\label{paramedcmintermsu}
\gamma_p=
\bar U^\dagger
\begin{pmatrix}
U^\star&0\\
0&U
\end{pmatrix}
\bar U
\tilde D
\bar U^T
\begin{pmatrix}
U^\dagger&0\\
0&U^T
\end{pmatrix}
\bar U^\star.
\end{equation}
Equation~\eqref{paramedcmintermsu} shows that the optimization in Eq.~\eqref{gsenergy} can equivalently be formulated over the positive squeezing parameters $\{d_i\}$ and the unitary matrix $U$. The latter can then be parameterized by a parametrized quantum circuit, providing a direct route to hybrid quantum--classical optimization~\cite{abbas2024challenges}.
\section{Symplectic Trace Minimization }\label{C1}
A majorization relation among symplectic eigenvalues, proven by Hiroshima~\cite{hiroshima2006additivity}, allows the symplectic diagonalization problem to be reformulated as a symplectic trace minimization problem~\cite{gao2021riemannian,son2021computing}:
\begin{equation}\label{sympstielf}
    \inf_{X \in \mathcal{S}_{p}(2k \times 2d ,\mathds{R})}
    \; \tfrac{1}{4}\operatorname{tr}(X H X^{T})
    = \frac{1}{2}\sum_{i=1}^{k} \epsilon^{\uparrow}_{i},
    \qquad 1 \le k \le d .
\end{equation}
Here $\epsilon^{\uparrow}_{i}$ denote the symplectic eigenvalues of $H$ ordered increasingly. In the main-text we demonstrated empirically that the choice
$X = P_{1} \mathcal{L}_{3} \in \mathcal{S}_{p}(2 \times 2d ,\mathds{R})$
accurately recovers the spectral gap via this variational principle.

More generally, we define the projector
\begin{equation}
    P_k =
    \begin{bmatrix}
        \mathds{1}_{k\times d} & 0_{k\times d} \\
        0_{k\times d} & \mathds{1}_{k\times d}
    \end{bmatrix}
    \in \mathbb{R}^{2k\times 2d},
\end{equation}
where $\mathds{1}_{k\times d}$ denotes the first $k$ rows of a $d\times d$ identity matrix and $0_{k\times d}$ the corresponding rows of the zero matrix.

We extend the methods of the main-text beyond the spectral gap and ground-state energy computation by considering the variational bound
\begin{equation}\label{sympsteilfunitri}
    \frac{1}{2}\sum_{i=1}^{k} \epsilon^{\uparrow}_{i}
    \;\leq\;
    \inf_{M_{1},M_{2},M_{3}}
    \frac{1}{4}\mathrm{tr}\!\left(
        P_{k}\mathcal{L}_{3}H\mathcal{L}^{T}_{3}P^{T}_{k}
    \right),
    \qquad 1 \le k \le d .
\end{equation}

\begin{figure}
    \centering
    \includegraphics{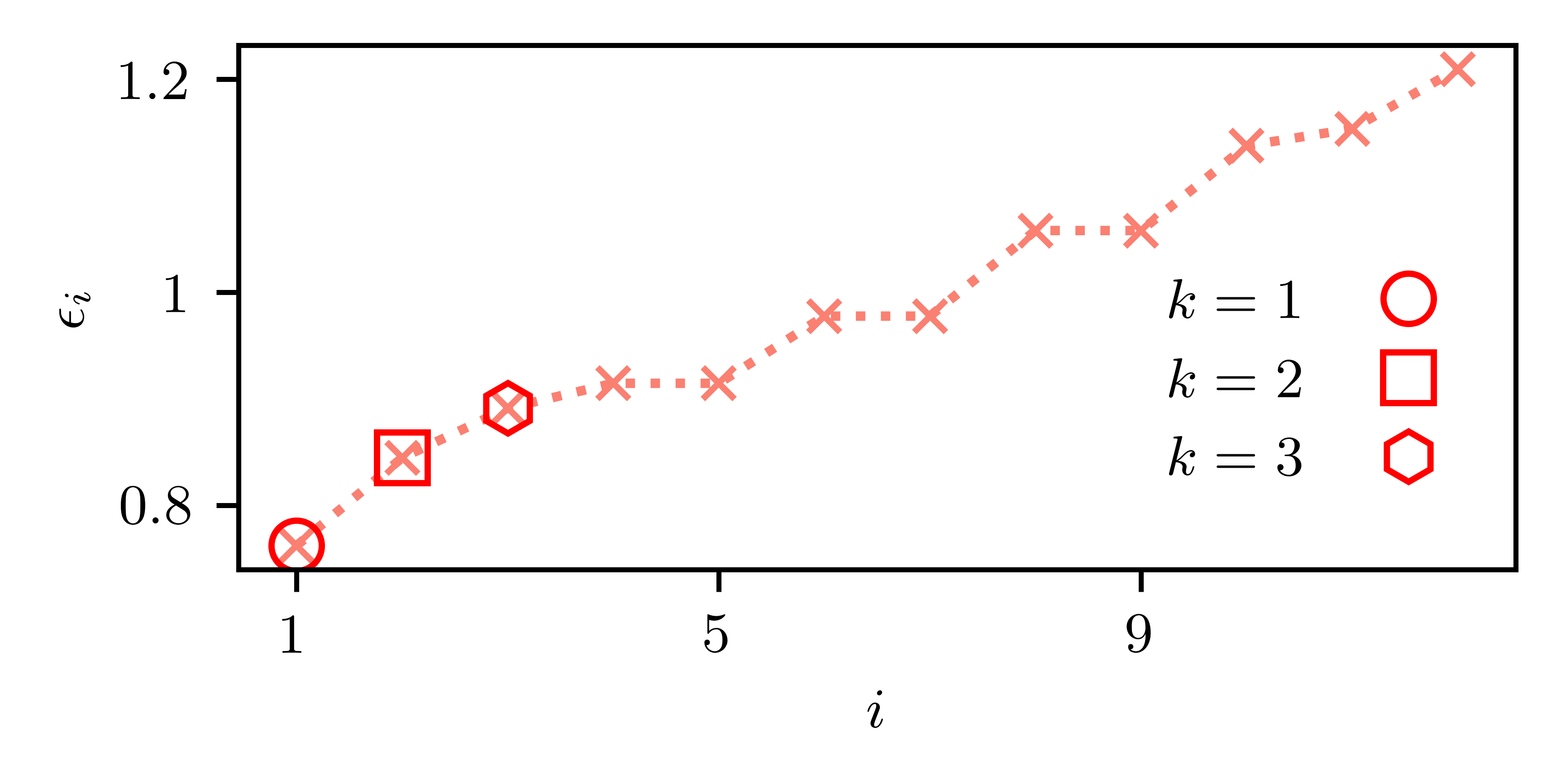}
    \caption{Utilizing the majorization relation in Eq. \eqref{sympstielf}, we show here that Sopt can retrieve the first three symplectic eigenvalues of a 12 mode symplectic spectrum, where we numerically solve the optimization in Eq. \eqref{sympsteilfunitri}, for different $k$ values. }
    \label{stefielsqr}
\end{figure}

To investigate the tightness of this bound, we consider the QDO system introduced in the main-text, arranged on a $2\times2$ square lattice (a 12-mode system), with $\rho=2$. We employ the conjugate gradient algorithm described previously to optimize the variational parameters in the matrices $M_{1}$, $M_{2}$, and $M_{3}$. We note that this parametrization is already overparameterized for the spectral gap computation and is therefore likely overparameterized for the estimation of higher symplectic eigenvalue sums as well; nevertheless, it provides a flexible variational ansatz with which to assess the saturation of the bound in practice. Fig.~\ref{stefielsqr} shows the agreement between the optimized symplectic spectrum and  exact symplectic diagonalization for the first three symplectic eigenvalues, $\epsilon_{1} \le \epsilon_{2} \le \epsilon_{3}$, with the residual discrepancy set by the solver tolerance.

\section{Non-Block Diagonal Hamiltonian}\label{D1}

\begin{figure*}
    \centering    \includegraphics{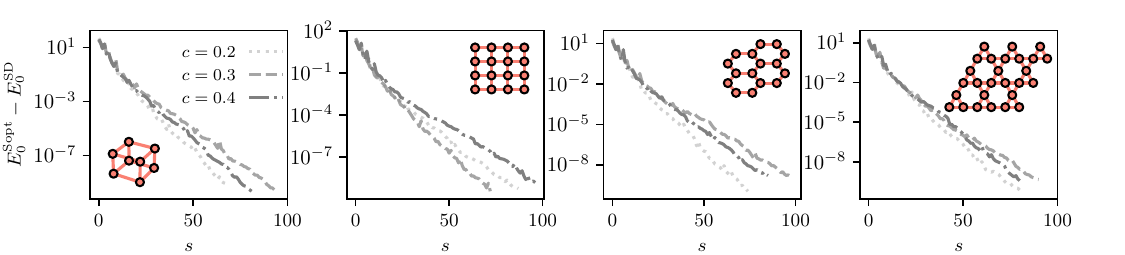}
    \caption{Sopt accurately retrieves ground-state energies on dipole coupled lattices, amended with diagonal position momentum coupling, with coupling strength given by $c$. All lattice calculations are shown at $\rho=1.9$, with $N=5\times 5\times 5$ in (a), $N=10\times 10$ in (b), $N=2\times 7\times 7$ in (c) and $N=3\times 5 \times 5$ in (d).   }
    \label{smfig2}
\end{figure*}

Here we extend the QDO model Hamiltonian considered in the main-text, to non-block diagonal couplings, with $2d\times 2d$ Hamiltonian matrix $H=\begin{pmatrix}
V & c \mathds{1}_{d} \\  c \mathds{1}_{d} &\mathds{1}_{d}
\end{pmatrix}$, with the strength of the diagonal position—momentum  coupling, given by $c \in \mathds{R}$. In Fig. \ref{smfig2}, we compare the ground-state energy computed via Sopt, with the ground-state energy computed via SD, using the same methods as outlined in the main-text.
This shows the generalization of the Sopt framework to quadratic bosonic Hamiltonians with coupling between position and momentum operators.  

\section{ Computing Gradients of Unit Triangular Cost Function}\label{E1}
Here we derive expressions for the gradients of the cost functions in Eq. \eqref{sympoptclasic} and Eq. \eqref{SoptDELTA} given in the main-text. 
Throughout this work, gradients are computed with respect to the Frobenius inner product on the unconstrained matrix parameters, corresponding to the Euclidean metric on the independent entries of the unit-triangular matrices.

The gradient of $\frac{1}{4}\mathrm{tr}(\mathcal{L}_{3} H\mathcal{L}_{3}^{T})$ with respect to $\mathcal{L}_{3}$ is \cite{petersen2008matrix},
\begin{equation}\label{gradl3}
  \frac{1}{4}\nabla_{\mathcal{L}_{3}}\mathrm{tr}(\mathcal{L}_{3} H\mathcal{L}_{3}^{T})=\frac{1}{2} \mathcal{L}_{3}H
 \end{equation}
Then using chain rule,  and the cyclic property of the trace operator, one can derive general expressions for the gradients of the variable matrices $M_{1}$, $M_{2}$ and $M_{3}$
\begin{equation}\label{m1}
\begin{aligned}
\frac{1}{4} \nabla_{M_{1}} \mathrm{tr}(\mathcal{L}_{3} H\mathcal{L}_{3}^{T} )
&= \Pi_{1}^{T}\mathcal{L}_{3}H\Pi_{2} M_{3}M_{2}
   + \Pi_{1}^{T}\mathcal{L}_{3}H\Pi_{2} \\
&\quad + \Pi_{1}^{T}\mathcal{L}_{3}H \Pi_{1}M_{2},
\end{aligned}
\end{equation}

\begin{equation}\label{m2}
\begin{aligned}
\frac{1}{4} \nabla_{M_{2}} \mathrm{tr}(\mathcal{L}_{3} H\mathcal{L}^{T}_{3})
&= M_{1}\Pi_{1}^{T}\mathcal{L}_{3}H\Pi_{2} M_{3}
   + M_{1}\Pi_{1}^{T}\mathcal{L}_{3}H\Pi_{1} \\
&\quad + \Pi_{2}^{T}\mathcal{L}_{3}H\Pi_{2}M_{3}
   + \Pi_{2}^{T}\mathcal{L}_{3}H\Pi_{1},
\end{aligned}
\end{equation}

\begin{equation}\label{m3}
\begin{aligned}
\frac{1}{4} \nabla_{M_{3}} \mathrm{tr}(\mathcal{L}_{3} H\mathcal{L}^{T}_{3})
&= M_{2}M_{1}\Pi_{1}^{T}\mathcal{L}_{3}H\Pi_{2}
   + \Pi_{1}^{T}\mathcal{L}_{3}H\Pi_{2} \\
&\quad + M_{2}\Pi_{2}^{T}\mathcal{L}_{3}H\Pi_{2}.
\end{aligned}
\end{equation}

Here $\Pi_{1}$ and $\Pi_{2}$ are the basis matrices defined in the main-text.
For the block-diagonal Hamiltonian of the form $V \oplus \mathds{1}_{d}$, we  write out the gradient expressions, in terms of  multiplications of $d \times d$ matrices, rather than $2d \times 2d$ sized matrix multiplications. We use the explicit form of $\mathcal{L}_{3}$,
\begin{equation}
\begin{aligned}
\mathcal{L}_{3}
&=\begin{bmatrix} \mathds{1}_{d} & M_{1}\\0 & \mathds{1}_{d} \end{bmatrix}
  \begin{bmatrix} \mathds{1}_{d}  & 0\\M_{2} & \mathds{1}_{d} \end{bmatrix}
  \begin{bmatrix} \mathds{1}_{d}  & M_{3}\\0 & \mathds{1}_{d} \end{bmatrix} \\
&=\begin{bmatrix}
        \mathds{1}_{d}+M_{1}M_{2} & M_{3}+M_{1}M_{2}M_{3}+M_{1}\\
        M_{2} & M_{2}M_{3}+\mathds{1}_{d}
  \end{bmatrix}.
\end{aligned}
\end{equation}

The gradients with respect to the variable matrices $M_{1}$, $M_{2}$ and $M_{3}$ are given as,
\begin{widetext}
\begin{equation}
\frac{1}{4}\,\nabla_{M_1}\, \mathrm{tr}\!\left(\mathcal{L}_3 V\oplus \mathds{1}_{d} \mathcal{L}^{T}_3\right)
= (M_{3}+M_{1}M_{2}M_{3}+M_{1})(\mathds{1}_{d}+M_{3}M_{2})+(\mathds{1}_{d}+M_{1}M_{2})V M_{2}
\end{equation}
\begin{equation}
\frac{1}{4}\,\nabla_{M_2}\, \mathrm{tr}\!\left(\mathcal{L}_3 V\oplus \mathds{1}_{d} \mathcal{L}^{T}_3\right)
= M_{1}((M_{3}+M_{1}M_{2}M_{3}+M_{1})M_{3}+(\mathds{1}_{d}+M_{1}M_{2})V)+(M_{2}M_{3}+\mathds{1}_{d})M_{3}+M_{2}V,
\end{equation}
\begin{equation}
\frac{1}{4}\,\nabla_{M_3}\, \mathrm{tr}\!\left(\mathcal{L}_3 V\oplus \mathds{1}_{d}\mathcal{L}_3^{T}\right)
= (\mathds{1}_{d}+M_{2}M_{1})(M_{3}+M_{1}M_{2}M_{3}+M_{1})+M_{2}(M_{2}M_{3}+\mathds{1}_{d}).
\end{equation}
\end{widetext}
We further include expressions for the gradient of Eq. \eqref{SoptDELTA} given in the main-text. Gradients are given for the objective $\frac{1}{4}\mathrm{tr}\!\left(
    P_{1}\mathcal{L}_{3}H\mathcal{L}^{T}_{3}P^{T}_{1}
    \right)$; the overall prefactor does not affect the minimizer. We define $A=(B^{T}P_{1})^{T}$ with $B=P_{1}\mathcal{L}_{3}H$. The definition of the projector $P_{1} \in \mathbb{R}^{2\times 2d}$ , given in the main-text, selects the first position and momentum component of $\mathcal{L}^{T}_{3}\mathcal{L}_{3}$. Selecting differently would require modification of the following gradient expressions.
Defining the basis vectors $e^{T}_{1}=\underbrace{[1,..,0,0,..,0]}_{2d}$ and $e^{T}_{2}=[0,..,0,1,..,0]$, we have the following expressions for the gradients of the variable vector $m_{1}$ and matrices $M_{2}$ and $M_{3}$,  
\begin{widetext}
\begin{equation}\label{m1_gap}
   \nabla_{m_1} \frac{1}{4}\mathrm{tr}(P_{1}\mathcal{L}_{3}H\mathcal{L}^{T}_{3}P^{T}_{1})=e^{T}_{1}A\Pi_{2}M_{3}M_{2}+e^{T}_{1}A\Pi_{2}+e^{T}_{1}A\Pi_{1}M_2,
\end{equation}
\begin{equation}\label{m2_gap}
   \nabla_{M_2} \frac{1}{4}\mathrm{tr}(P_{1}\mathcal{L}_{3}H\mathcal{L}^{T}_{3}P^{T}_{1})=m_1 \otimes (e^{T}_{1}A\Pi_{2} M_{3}+e^{T}_{1}A\Pi_{1})+\underbrace{(1,0,...,0)}_{d} \otimes (e^{T}_{2}A\Pi_{1}+e^{T}_{2}A\Pi_{2}M_{3})
\end{equation}
and 
\begin{equation}\label{m3_gap}
   \nabla_{M_3} \frac{1}{4}\mathrm{tr}(P_{1}\mathcal{L}_{3}H\mathcal{L}^{T}_{3}P^{T}_{1})=M_{2}m_{1}\otimes e^{T}_{1}A\Pi_{2}+\underbrace{(1,0,...,0)}_{d}M_{2} \otimes e^{T}_{2}A\Pi_{2}+\underbrace{(1,0,...,0)}_{d}\otimes e^{T}_{1}A\Pi_{2}.
\end{equation}
\end{widetext}
Here $\otimes $ denotes the outer product. Note that whether the number of variational parameters in Eq. \eqref{SoptDELTA} is optimal is not determined here, therefore the expressions in Eq. \eqref{m1_gap} –Eq. \eqref{m3_gap} may include redundant terms. 

In the numerical implementation, the symmetric matrices $M_k$ with $k \in\{1,2,3\}$ are parameterized via unconstrained upper-triangular matrices $X_k$ as $M_k = \operatorname{triu}(X_k) + \operatorname{triu}(X_k)^{T}$.
Gradients are therefore evaluated with respect to $X_k$ and projected onto the symmetric subspace, which absorbs the factor $1/2$ appearing in Eq.~(\ref{gradl3}).

\section{Additional numerical details}\label{F1}
The convergence study shown in the inset of
Fig.~\ref{sweeps}(a) was performed using uniformly spaced
$\rho$-sweeps with
\begin{center}
\small
\begin{tabular}{l}
$n_\rho=$
25, 50, 75, 100, 125, 150, 175, 200, 225,\\
250, 275, 300, 350, 400, 500, 800, 1000.
\end{tabular}
\end{center}
\section{Banded warm-started parameterization}\label{H}
\begin{figure}
    \centering
    \includegraphics{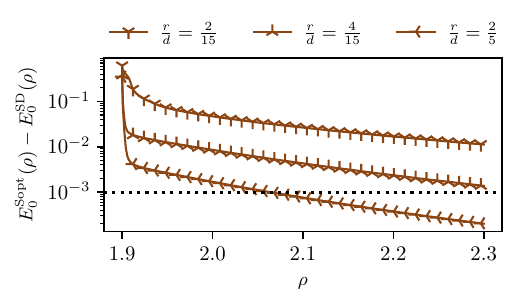}
    \caption{
Warm-started $\rho$-sweeps using symmetric banded parameter matrices of bandwidth $r$ for QDOs on a $5\times5$ square lattice. The matrices $M_1$, $M_2$, and $M_3$ are all restricted to the same bandwidth, and 4 optimization steps are performed at each $\rho$ point using the warm-starting procedure described in Sec \ref{sweepsec}. The three curves correspond to bandwidths $r/d=2/15$, $4/15$, and $2/5$, corresponding to approximately $87\%$, $73\%$, and $60\%$ reductions in the number of variational parameters, respectively.
}
    \label{bandedplot}
\end{figure}

Physically motivated parameter reductions provide a natural direction for reducing the computational cost and memory requirements of Sopt. We restrict the variable matrices $M_1$, $M_2$ and $M_3$ to symmetric banded matrices with bandwidth $r$, while retaining the warm-started sweep procedure of Sec \ref{sweepsec}.
\section{Riemannian Cayley-retraction method}\label{G1}
We implement the Riemannian Cayley-retraction method, applying the approach detailed in Ref. \cite{son2021computing}. The initial symplectic matrix, given as input to the Riemannian symplectic trace minimization procedure  is $X_{1}=\gamma_{\mathcal{T}}$ (see main-text for definition), as is the case for the Sopt initialization. The Riemannian gradient of the trace cost function at $X_{i}$, where $X_{i} \in \mathcal{S}_{p}(2d,\mathds{R}) $ is given as, $ \mathrm{grad}^{p}_{X_{i}}=\bar{A}^{p}_{X_{i}}\sigma X_{i}$ with $\bar{A}^{p}_{X_{i}}=\frac{1}{2}\left(A^{p}_{X_{i}}+(A^{p}_{X_{i}})^{T}\right)$, where $A^{p}_{X_{i}}=4H^{p}_{X_{i}}HX_{i}(X_{i}\sigma)^{T}$ and $H^{p}_{X_{i}}=\mathds{1}_{2d}+\frac{p}{2}X_{i}X_{i}^{T}-\sigma X_{i}(X_{i}^{T}X_{i})^{-1}X_{i}^{T}\sigma^{T}$, where recall $\sigma$ is given in Eq. \eqref{sigmasymplectic} and $H$ is the Hamiltonian matrix, given in Eq. \eqref{bosonicquad}.
The Cayley-retraction step, uses the Riemannian gradient to compute the next symplectic matrix $X_{i+1} \in \mathcal{S}_{p}(2d,\mathds{R})$, via the update rule:
\begin{equation}\label{cayleystep}
    X_{i+1}= \left(\mathds{1}+\frac{t}{2}\bar{A}^{p}_{X_{i}}\sigma\right)^{-1}\left(\mathds{1}-\frac{t}{2}\bar{A}^{p}_{X_{i}}\sigma\right)X_{i},
\end{equation}
where here $t,p \in \mathds{R}$, with $t=0.1$ and $p=1$ for the examples shown in the main-text. In the timing implementation, the Cayley retraction in Eq. \eqref{cayleystep} was evaluated
using a linear solve with multiple right-hand sides rather than by explicitly forming the matrix
inverse. The Riemannian gradient construction also involves inverse
matrix factors; these were evaluated directly in the present reference
implementation.
\bibliography{ref2}
\end{document}